\documentclass[fleqn,usenatbib]{mnras}
\usepackage{amsmath,amssymb,txfonts}
\usepackage{graphicx,color}
\usepackage{threeparttable}
\usepackage{url}

\newcommand{\ergs}{{\rm \,erg\,s^{-1}}}
\newcommand{\msun}{M_{\sun}}
\newcommand{\mbh}{{M_{\rm BH}}}
\newcommand{\ledd}{{L_{\rm Edd}}}
\newcommand{\lbol}{{L_{\rm bol}}}
\newcommand{\lx}{{L_{\rm X}}}
\newcommand{\lhx}{{L_{\rm HX}}}
\newcommand{\lr}{{L_{\rm R}}}
\newcommand{\xix}{{\xi_{\rm X}}}
\newcommand{\xim}{{\xi_{\rm M}}}
\newcommand{\pjet}{{P_{\rm jet}}}
\newcommand{\njet}{{\eta_{\rm jet}}}
\newcommand{\fx}{{f_{\rm X}}}
\newcommand{\lbolh}{{L_{\rm bol,h}}}

\newcommand{\mk}{{\rm M}_{K}}

\title[Universal RX Correlation in Seyferts]{Possible Evidence of a Universal Radio/X-ray Correlation in a near-Complete Sample of Hard X-ray Selected Seyfert Galaxies}
\author[N. Chang et al.]{N. Chang$^{1, 2}$\thanks{E-mail: changning@xao.ac.cn (NC), fgxie@shao.ac.cn (FGX), liux@xao.ac.cn (XL)}, F.~G. Xie$^{3\star}$, X. Liu$^{1, 4\star}$, L.~C. Ho$^{5, 6}$, A.-J. Dong$^{1, 7}$, Z.~H. Han$^8$, X. Wang$^{1, 2}$\\
$^1$ Xinjiang Astronomical Observatory, Chinese Academy of Sciences, 150 Science 1-Street, Urumqi 830011, Xinjiang, China\\
$^2$ School of Astronomy and Space Science, University of Chinese Academy of Sciences, Beijing, 100049, China\\
$^3$ Key Laboratory for Research in Galaxies and Cosmology, Shanghai Astronomical Observatory, Chinese Academy of Sciences, 80 Nandan Road, \\
Shanghai 200030, China\\
$^4$ Key Laboratory of Radio Astronomy, Chinese Academy of Sciences, Nanjing 210008, China\\
$^5$ Kavli Institute for Astronomy and Astrophysics, Peking University, Beijing 100871, China\\
$^6$ Department of Astronomy, School of Physics, Peking University, Beijing 100871, China\\
$^7$ Guizhou Provincial Key Laboratory of Radio Astronomy and Data Processing, Guizhou Normal University, Guiyang 550001, China\\
$^8$ Physics and Electronic Engineering Department, Xinjiang Normal University, Urumqi 830000, China}

\date{Accepted XXX. Received YYY; in original form ZZZ}

\begin{document}
\defcitealias{Panessa2015}{P15}

\label{firstpage}
\pagerange{\pageref{firstpage}--\pageref{lastpage}}
\maketitle

\begin{abstract}
Because the disc--jet coupling likely depends on various properties of sources probed, the sample control is always an important but challenging task. In this work, we re-analyzed the {\it INTEGRAL} hard X-ray-selected sample of Seyfert galaxies. We only consider sources that have measurements in black hole mass, and luminosities in radio and X-rays. Our final sample includes 64 (out of the original 79) sources, consists of both bright AGNs and low-luminosity ones. The 2-10 keV X-ray Eddington ratio $\lx/\ledd$ locates in the range between $\sim 10^{-4.5}$ and $\sim 10^{-0.5}$. We first find that, because of the similarity in the $\lhx/\lx$ distribution, the X-ray origin of radio-loud Seyferts may be the same to that of radio-quiet ones, where we attribute to the hot accretion flow (or similarly, the corona). We then investigate the connections between luminosities in radio and X-rays. Since our sample suffers a selection bias of a black hole mass $M_{\rm BH}$ dependence on $\lx/\ledd$, we focus on the correlation slope $\xi_{\rm X}$ between the radio (at 1.4 GHz) and X-ray luminosities in Eddington unit, i.e. $(\lr/\ledd) \propto (\lx/\ledd)^{\xi_{\rm X}}$. We classify the sources according to various properties, i.e. 1) Seyfert classification, 2) radio loudness, and 3) radio morphology. We find that, despite these differences in classification, all the sources in our sample are consistent with a universal correlation slope $\xi_{\rm X}$ (note that the normalization may be different), with $\xi_{\rm X}=0.77\pm0.10$. This is unexpected, considering various possible radio emitters in radio-quiet systems. For the jet (either relativistic and well collimated, or sub-relativistic and weakly collimated) interpretation, our result may suggest a common/universal but to be identified jet launching mechanism among all the Seyfert galaxies, while properties like black hole spin and magnetic field strength only play secondary roles. We further estimate the jet production efficiency $\eta_{\rm jet}$ of Seyfert galaxies, which is $\eta_{\rm jet}\approx1.9^{+0.9}_{-1.5}\times10^{-4}$ on average. We also find that $\eta_{\rm jet}$ increases as the system goes fainter. Alternative scenarios for the radio emission in radio-quiet systems are also discussed.
\end{abstract}

\begin{keywords}
galaxies: active -- galaxies: Seyfert -- radio continuum: galaxies -- accretion: accretion discs
\end{keywords}

\section{Introduction}\label{sec:intro}

Almost every galaxy contains a supermassive black hole (BH) at its center (\citealt{KormendyHo2013}). Among most of the cosmic time the BH remains quiet; only over a short period of $10^4 - 10^6$ yrs (e.g., \citealt*{KauffmannHaechnelt2000,MartiniWeinberg2001,Condon2013}), the BH is fed with a sufficient amount of gas/material and the system becomes active and luminous, i.e. enter a phase so-called the active galactic nucleus (AGN). It is now widely accepted that the AGN and its host galaxy are tightly connected, i.e. the host galaxy provides the feed for the accretion, while the energy and momentum driven by the AGN will also impact the dynamics of the gas (consequently, the star formation) within the galaxy (for reviews, see e.g., \citealt*{Fabian2012,KormendyHo2013,HeckmanBest2014}). 

The structure of AGN is admittedly complex. Emission at different wavebands may originate from different physical components or different spatial locations (see e.g., \citealt{Ho2008,HeckmanBest2014} for reviews), i.e. continuum radiation in radio, optical-ultraviolet (UV), and hard X-rays originates from, respectively, the relativistic jet, the cold accretion disc and the corona (or hot accretion flow in case of low-luminosity AGNs, here we use them interchangeably). The highly collimated jet, which propagates from sub-pc up to Mpc scales, may play a crucial role in the BH-host galaxy co-evolution (e.g., \citealt{Croton2006,  HeckmanBest2014}), although it remains inclusive on its basic physics, e.g., its composition and the launching and acceleration/deceleration mechanisms (see e.g., \citealt{HarrisKrawczynski2006, Tchekhovskoy2015, DavisTchekhovskoy2020} for summaries, and \citealt{ParfreyPhilippovCerutti2019, ChenZhang2020} for recent progress). Statistically, as the systems become fainter, the fraction of radio-loud AGNs increases \citep{Ho2008}, which agrees with observations in black hole binaries \citep{Fender2009}. Observationally jets in AGNs are diverse in both power and morphology \citep{Padovani2016, Panessa2019, chiaraluce2020}. Depending on the radio-to-optical luminosity ratio, the AGNs can be separated into two main classes. The minorities are radio-loud (RL), where the jet is highly relativistic and well collimated (for reviews, see e.g., \citealt{HarrisKrawczynski2006, BlandfordMeierReadhead2019, HardcastleCroston2020}). Based on their large-scale radio morphology, the RL AGNs can be further classified to FR I or II \citep{FanaroffRiley1974}. The rest majorities are radio-quiet (RQ), where the radio emission is compact, without clear well-collimated structure in high-resolution observations. The origin of the radio emission in RQ AGNs is still under debate, a significant fraction of the large-scale radio emission may originate from stellar processes, or from interactions between AGN wind and interstellar medium (e.g., \citealt*{Condon1992, Padovani2015, Panessa2019, chiaraluce2020}). Even for the compact radio core (of RQ systems) investigated in this work, the origin is still inclusive, either a weak jet adopted here, or nuclear activities related to star formation (e.g., \citealt*{LaorBehar2008, Bonchi2013, Baek2019, Smith2020}). Under the jet interpretation, the jet velocity is low (e.g., non-relativistic) and the collimation of jet is poor. Despite these observational differences, it is argued that jet among these systems may be governed by the same physics \citep{ChenZhang2020}.

For the investigation of the connection of the jet and the accretion flow (more clearly, the hot accretion flow. See \citealt{YuanNarayan2014} for a review) in AGNs and BH X-ray binaries (BHBs), a tight linear relationship in logarithmic space among the BH mass $\mbh$, and the radio (monochromatic, $\lr=\nu L_\nu$ at e.g. 1.4 GHz) and X-ray (integrated, i.e. $\lx = \int L_\nu d\nu$ in e.g. the 2--10 keV band) luminosities has been discovered (e.g., \citealt*{Merloni2003, Falcke2004, Panessa2007,  LiWuWang2008, Gultekin2009, Gultekin2019, Burlon2013, Liu2016, Inoue2017, Qian2018}). It is also called the ''fundamental plane'' (FP) of BH activity.\footnote{Note that there are at least two types of jets, one is continuous/steady and the other is transient/episodic shown as discrete blobs, see e.g., \citet*{Fender2009} for the classification of these two types of jets in BHBs. There may also exist a third type of jet (e.g., \citealt*{XieYanWu2020, Zdziarski2020} and references therein). In the FP studies only the continuous/steady jets are considered.} Motivated by accretion theory (see this suggestion in \citealt{XieYuan2017}), we consider the FP in a revised space, i.e. ($\log(\lr/\ledd), \log(\lx/\ledd), \log\mbh$) and re-express it as,
\begin{equation}
\log(\lr/\ledd) = \xix \log(\lx/\ledd) + \xim \log\mbh + const.. \label{eq_fp}
\end{equation}
Here $\ledd = 1.26\times10^{46}\,(\mbh/10^8 \msun)\,\ergs$ is the Eddington luminosity. $\lr/\ledd$ and $\lx/\ledd$ define the Eddington ratios in radio and  X-rays, respectively. We focus on the dependence on the luminosity (i.e. slope parameter $\xix$), but not on the BH mass (i.e. $\xim$). It is found that a majority of sources follow a ``standard'' $\xix\approx 0.6\pm0.1$ FP (see e.g., \citealt*{Merloni2003, Gultekin2009, Gultekin2019, Qian2018}, and for the case of BHBs only, see e.g., \citealt*{Corbel2003,Corbel2013}). Studies suggest that the FP exist only in less luminous (e.g., sub-Eddington, see \citealt{MerloniHeinz2008}) systems. Theoretically the FP can be naturally interpreted under the framework of an accretion--jet model (e.g., \citealt*{HeinzSunyaev2003, YuanCuiNarayan2005, XieYuan2016}), where a hot accretion flow co-exists with a relativistic or mildly relativistic jet.\footnote{As stated above, the origin of radio emission in RQ AGNs is questioned in recent years (e.g., \citealt*{Bonchi2013, Baek2019, Smith2020}, see \citealt*{Panessa2019} for a review), where they argue the contributions from nuclear star formation, outflow, corona and jet are of comparable importance in many RQ AGNs.} Observationally, individual sources may have a large scatter to the FP. Such scatter can be due to effects of, among others, the (combination of) BH spin \citep{Miller2009,UnalLoeb2020}, the strength of the magnetic field \citep{BlandfordZnajek1977,Sikora2007,LiXie2017}, the Doppler beaming effect \citep{LiWuWang2008}, and the environment \citep{vanVelzenFalcke2013}.

Ever since its discovery, deviations to the standard FP in $\xix$ are observed in different classes of systems, e.g., the radio-loud AGNs (e.g., \citealt*{WangWuKong2006,Panessa2007,LiWuWang2008,deGasperin2011}), and the narrow-line Seyfert 1 galaxies (e.g., \citealt{Yao2018}). Such deviation is also observed in BHBs (e.g., \citealt*{Coriat2011,Corbel2013,XieYanWu2020}). It is thus clear that $\xix$ depends on the sample compilation, and its value may hint on the accretion mode of individual (type) of sources (e.g., \citealt*{HeinzSunyaev2003,XieYuan2016, XieYanWu2020}). Indeed, theoretically we can link the radio and X-ray luminosities to the dimensionless mass accretion rate $\dot{m}$\footnote{Dimensionless accretion rate $\dot{m}$ is defined as $\dot{m} = \dot{M}/\dot{M}_{\rm Edd}$, i.e. the mass accretion rate $\dot{M}$ normalized by the Eddington accretion rate $\dot{M}_{\rm Edd} = 10\,\ledd/c^2$.} as (see e.g., \citealt*{UnalLoeb2020})
\begin{equation}
\lr/\ledd \propto \dot{m}^{\gamma}, \hspace{0.3cm}{\rm and}\hspace{0.3cm} \lx/\ledd \propto \dot{m}^{\kappa},
 \end{equation}
where $\gamma\approx 1.3-1.4$ (e.g., \citealt{HeinzSunyaev2003}). In the above expressions, additional dependence on, e.g., magnetic field strength and BH spin, is omitted for simplicity. Parameters $\gamma$ and $\kappa$ characterize respectively, the radiative efficiencies of jet and hot accretion flow, i.e. $\lr/(\dot{M} c^2) \propto \dot{m}^{\gamma-1}$ and $\lx/(\dot{M} c^2) \propto \dot{m}^{\kappa-1}$. Physically the observation of $\xix$ then measures the value of $\gamma/\kappa\approx(1.3-1.4)/\kappa$. If we additionally estimate the jet power as $P_{\rm jet}\propto L_{\rm R}^{6/7}$ \citep{Willott1999, Cavagnolo2010, Su2017}, then the FP also provides information on the connection between powers in accretion and ejection, which can then used to probe the efficiency of converting accretion power into ejection (e.g., \citealt{Inoue2017, Rusinek2020, SoaresNemmen2020, Wojtowicz2020}).

\begin{figure*}
  \vspace{-0.3cm}
  \includegraphics[width=8cm]{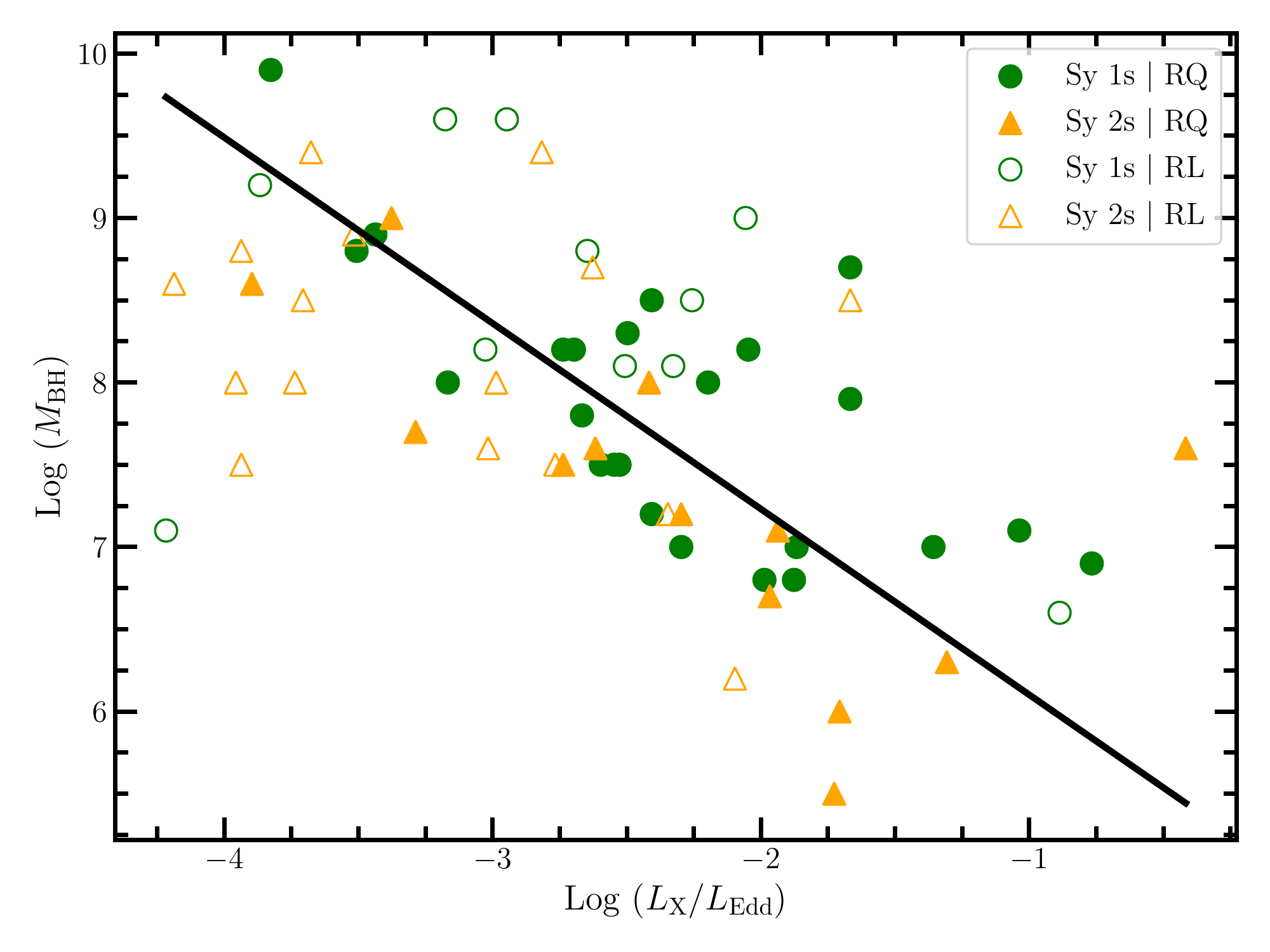}\hspace{0.3cm}
  \includegraphics[width=8cm]{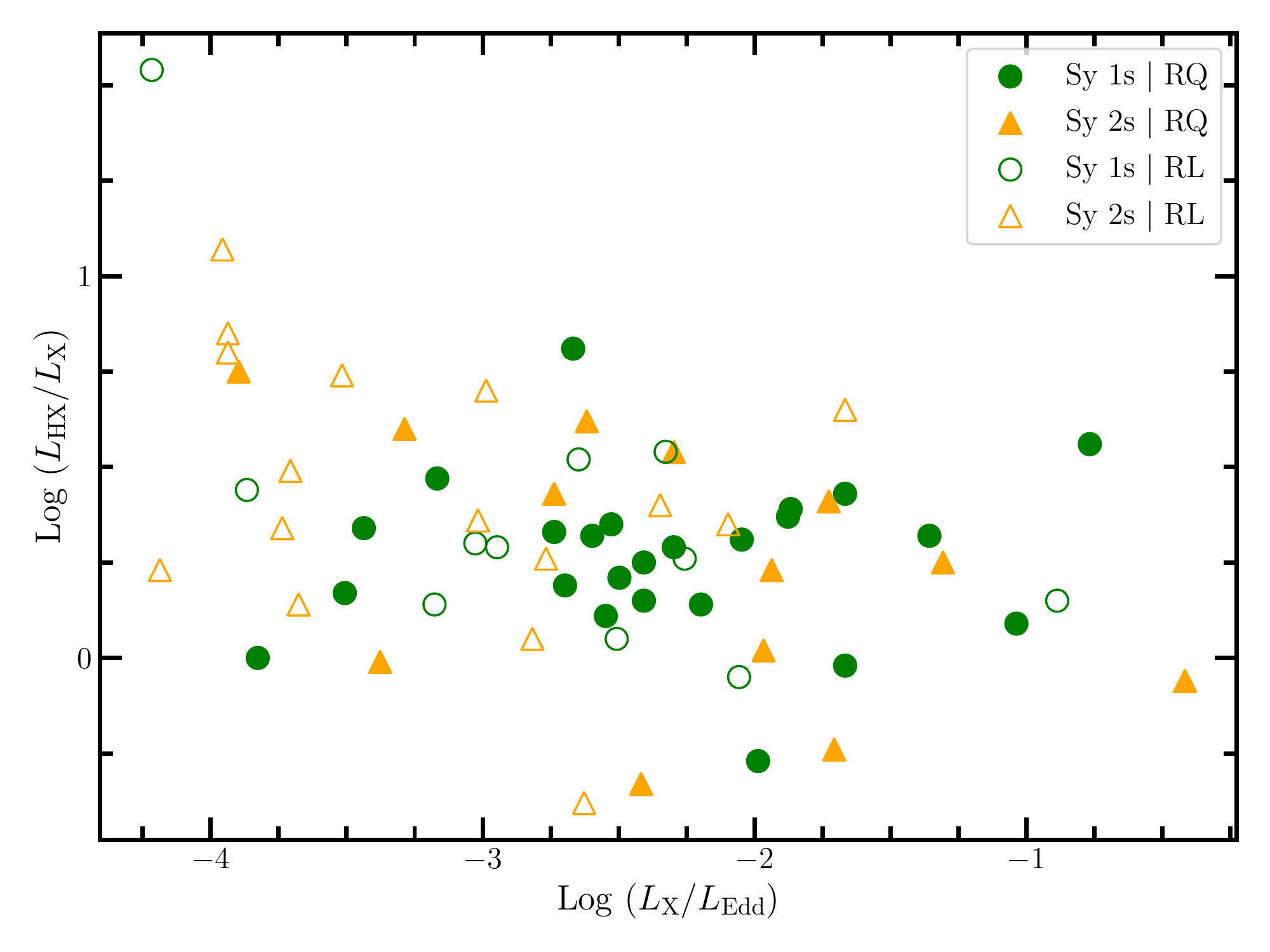}
  \caption{Distributions of BH mass $\mbh$ (left panel) and the 20-100 keV to 2-10 keV X-ray flux ratio $\lhx/\lx$ (right panel) versus 2-10 keV X-ray Eddington ratio $\lx/\ledd$. In both panels, the open and filled separate sources in their radio loudness, i.e. open for radio-loud (RL) and filled for radio-quiet (RQ); while the color and shape define their optical properties, i.e. green circles for Seyfert 1s (labelled Sy 1s) and orange triangles for Seyfert 2s (labelled Sy 2s). The solid curve in the left panel shows a simple power-law fit with $\mbh\propto (\lx/\ledd)^{-1.13\pm0.15}$.}
  \label{fig1} 
\end{figure*}

With the advantage of free from absorption, the  $\ga 10$ keV hard X-rays provide a more direct measurement of the accretion power (especially of the hot accretion flow, which tightly correlates with jet). An unbiased hard X-ray selected sample, e.g., the {\it INTEGRAL}/IBIS selected sample of \citealt{Panessa2015} (hereafter \citetalias{Panessa2015}; see Section \ref{sec:sample}), will then be of great importance in the disc--jet coupling investigation. Most sources in the \citetalias{Panessa2015} Seyfert sample have $\lx/\ledd \ga 10^{-3}$, i.e. they belong to bright AGNs, and only a limited number of sources belong to low-luminosity ones ($\lx/\ledd \la 10^{-3}$). It is known that bright AGNs and low-luminosity ones are distinctive in their accretion physics \citep{Ho2008, HeckmanBest2014}. Theoretically bright AGNs are powered by cold accretion disk (e.g., Shakura-Sunyaev disc [SSD], \citealt*{ShakuraSunyaev1973}), while low-luminosity AGNs are powered by hot accretion flow \citep{YuanNarayan2014}.

\citetalias{Panessa2015} focused on correlations among luminosities of their absolute values (e.g., $\lr$ and $\lx$), but not the Eddington normalized ones (e.g., $\lr/\ledd$ and $\lx/\ledd$), thus differences in accretion physics are not directly demonstrated. In this work, we reanalyze the complete hard-X-ray selected Seyfert galaxy sample of \citetalias{Panessa2015}, with an emphasis on comparison between different type of sources (radio loud versus radio quiet, Seyfert 1s versus Seyfert 2s, etc). For tens of sources we update their measurements in black hole mass and radio luminosity. This work is organized as follows. In Section \ref{sec:sample_stat}, we briefly introduce the sample, with a focus on those updates compared to \citetalias{Panessa2015}. The statistical method is also introduced here. Then in Section \ref{sec:result} we present our results in detail, where we compare properties among different subsamples. Section \ref{sec:discuss} is devoted to discussions of our results and the final Section \ref{sec:summary} provides a brief summary. Throughout this work, the distances are derived based on a flat cold dark matter cosmology ($\Lambda$CDM cosmology) with the Hubble constant $H_0= 70\, {\rm km\,s^{-1}\,Mpc^{-1}}$ and the mass density fraction parameter $\Omega_{\rm M}=0.3$.

\begin{table*}
\centering
\caption{Basic properties of the {\it INTEGRAL}/IBIS hard-X-ray selected Seyfert sample} \label{tab1}
\begin{threeparttable} 
\scalebox{0.8}{
\begin{tabular}{lcccccccccc}
\hline
Name & Class & $z$ & Log($\mbh$) & $\mbh$ Method & Ref. & Log($\lx$) & Log($\lhx$) & Log($\lr$) & Radio Mor.$^\star$ & $D_{\rm maj}$ \\
(1) & (2) & (3) & (4) & (5) & (6) & (7) & (8) & (9) & (10) & (11) \\
\hline
IGR J00333+6122 & Sy 1.5 & 0.105 & 8.5 & GR & 1 & 44.19 & 44.44 & 39.46 & U & $<$ 49.09 \\
NGC 788 & Sy 2 & 0.0136 & 7.5 & WL & 1 & 42.86 & 43.29 & 37.61 & R & 35.6 \\
NGC 1068 & Sy 2 & 0.0038 & 7.2 & WL & 1 & 42.95 & 43.35 & 39.27 & R & 18.95 \\
QSO B0241+62 & Sy 1 & 0.044 & 8.1 & WL & 4 & 43.87 & 44.41 & 40.3 & U & $<$ 20.57 \\
NGC 1142 & Sy 2 & 0.0288 & 9.4 & WL & 1 & 43.82 & 43.96 & 39.53 & U & $<$ 13.46 \\
B3 0309+411 & Sy 1 & 0.136 & 8.8 & $\mk$ & 2 & 45.04 & 44.99 & 41.27 & R & 1373 \\ 
NGC 1275 & Sy 2 & 0.0175 & 8.5 & WL & 1 & 42.89 & 43.38 & 39.60 (a) & R & 109.1 \\
3C 111 & Sy 1 & 0.0485 & 9.6 & WL & 1 & 44.52 & 44.66 & 41.38 & R & 393.0 \\
LEDA 168563 & Sy 1 & 0.029 & 8.0 & $\mk$ & 1 & 43.9 & 44.04 & 38.48 & S & 15.68 \\
4U 0517+17 & Sy 1.5 & 0.0179 & 7.0 & RM & 1 & 43.23 & 43.62 & 37.75 & S & 8.484 \\
MCG+08-11-11 & Sy 1.5 & 0.0205 & 8.1 & WL & 1 & 43.69 & 43.74 & 39.45 & R & 84.34 \\
Mrk 3 & Sy 2 & 0.0135 & 8.7 & WL & 1 & 44.17 & 43.79 & 39.76 & U & $<$ 6.311 \\
Mrk 6 & Sy 1.5 & 0.0188 & 8.2 & WL & 1 & 43.27 & 43.57 & 39.43 & U & $<$ 8.789 \\
IGR J07565-4139 & Sy 2 & 0.021 & 8.0 & $\mk$ & 3 & 42.14 & 43.21 & 37.84 & S$^*$ & 14.05 \\
IGR J07597-3842 & Sy 1.2 & 0.04 & 8.3 & GR & 1 & 43.9 & 44.11 & 38.24 & U & $<$ 18.7 \\
ESO 209-12 & Sy 1 & 0.0396 & 9.0 & $\mk$ & 2 & 43.43 & 43.87 & 29.23 & S$^*$ & 36.7 \\
FRL 1146 & Sy 1.5 & 0.0316 & 8.7 & $\mk$ & 2 & 43.39 & 43.56 & 38.61 & R & 94.55 \\
SWIFT J0917.2-6221 & Sy 1 & 0.0573 & 9.9 & G & 1 & 44.17 & 44.17 & 39.6 & S$^*$ & 35.35 \\
MCG-05-23-16 & Sy 2 & 0.0085 & 6.3 & WL & 1 & 43.09 & 43.34 & 37.47 & U & $<$ 3.974 \\
IGR J09523-6231 & Sy 1.9 & 0.252 & 7.6 & WL & 1 & 45.28 & 45.22 & 40.02 & U$^*$ & $<$ 112.6 \\
SWIFT  J1009.3-4250 & Sy 2 & 0.033 & 8.8 & $\mk$ & 2 & 42.96 & 43.81 & 38.74 & S$^*$ & 28.52 \\
NGC 3281 & Sy 2 & 0.0115 & 8.0 & WL & 1 & 43.68 & 43.35 & 38.45 & U & $<$ 5.376 \\
IGR J10404-4625 & Sy 2 & 0.2392 & 8.5 & S & 4 & 44.93 & 45.58 & 40.84 & S$^*$ & 204 \\
NGC 3783 & Sy 1 & 0.0097 & 7.5 & WL & 1 & 43.07 & 43.42 & 37.24 (b) & R & 55.63 \\
IGR J12026-5349 & Sy 2 & 0.028 & 8.0 & S & 4 & 43.11 & 43.81 & 39.06 & R & 143.1 \\
NGC 4151 & Sy 1.5 & 0.0033 & 7.5 & WL & 1 & 43.05 & 43.16 & 37.07 (c) & U & $<$ 1.543 \\
NGC 4388 & Sy 2 & 0.0084 & 7.2 & WL & 1 & 43.0 & 43.54 & 36.58 (d) & R & 49.22 \\
NGC 4507 & Sy 2 & 0.0118 & 7.6 & WL & 1 & 43.08 & 43.7 & 38.28 & S & 8.086 \\
LEDA 170194 & Sy 2 & 0.036 & 8.9 & WL & 1 & 43.48 & 44.22 & 39.13 & R & 130.2 \\
NGC 4593 & Sy 1 & 0.009 & 7.0 & WL & 1 & 42.8 & 43.09 & 37.18 & R & 35.9 \\
IGR J12415-5750 & Sy 1 & 0.023 & 8.0 & GR & 1 & 42.93 & 43.4 & 38.37 & S$^*$ & 15.83 \\
NGC 4945 & Sy 2 & 0.0019 & 6.2 & WL & 1 & 42.2 & 42.55 & 38.62 & R & 27.0 \\
Cen A & Sy 2 & 0.0018 & 8.0 & WL & 1 & 42.36 & 42.7 & 39.35 & R$^*$ & 33.66 \\
NGC 5252 & Sy 2 & 0.023 & 9.0 & WL & 1 & 43.72 & 43.71 & 38.39 & U & $<$ 10.75 \\
IC 4329A & Sy 1.2 & 0.016 & 7.0 & WL & 1 & 43.74 & 44.06 & 38.67 & R & 113.7 \\
Circinus Galaxy & Sy 2 & 0.0014 & 6.0 & WL & 1 & 42.39 & 42.15 & 37.64 & R$^*$ & 16.41 \\
NGC 5506 & Sy 1.9 & 0.0062 & 6.7 & WL & 1 & 42.83 & 42.85 & 37.82 (e) & U & $<$ 2.898 \\
ESO 511-G030 & Sy 1 & 0.2239 & 8.7 & $\mk$ & 1 & 45.13 & 45.56 & 40.08 & R & 224.8 \\
IGR J14515-5542 & Sy 2 & 0.018 & 7.6 & WL & 4 & 42.68 & 43.04 & 38.19 & S$^*$ & 11.79 \\
IC 4518A & Sy 2 & 0.0163 & 7.5 & G & 1 & 42.83 & 43.09 & 39.05 & R$^*$ & 71.12 \\
IGR J16024-6107 & Sy 2 & 0.011 & 7.4 & $\mk$ & 2 & 41.66 & 42.46 & 37.49 & U$^*$ & $<$ 4.914 \\
IGR J16351-5806 & Sy 2 & 0.0091 & 8.6 & $\mk$ & 2 & 42.51 & 42.74 & 38.22 & R$^*$ & 30.63 \\
IGR J16385-2057 & NLS1 & 0.0269 & 6.8 & GR & 1 & 43.02 & 43.39 & 38.13 & S & 15.34 \\
IGR J16482-3036 & Sy 1 & 0.031 & 8.2 & GR & 1 & 43.6 & 43.79 & 37.89 & R & 112.1 \\
IGR J16558-5203 & Sy 1.2 & 0.054 & 7.9 & GR & 1 & 44.33 & 44.31 & 38.65 & U$^*$ & $<$ 24.12 \\
NGC 6300 & Sy 2 & 0.0037 & 5.5 & WL & 1 & 41.87 & 42.28 & 36.92 & R$^*$ & 20.3 \\
GRS 1734-292 & Sy 1 & 0.0214 & 8.9 & WL & 1 & 43.56 & 43.9 & 38.79 & U & 10.0 \\
2E 1739.1-1210 & Sy 1 & 0.037 & 8.2 & GR & 1 & 43.56 & 43.89 & 38.13 & U & 17.3 \\
IGR J18027-1455 & Sy 1 & 0.035 & 7.6 & WL & 4 & 43.23 & 44.04 & 38.51 & R & 144 \\
ESO 103-35 & Sy 2 & 0.0133 & 7.1 & WL & 1 & 43.26 & 43.49 & 38.1 & U$^*$ & $<$ 5.941 \\
3C 390.3 & Sy 1 & 0.0561 & 8.5 & WL & 1 & 44.34 & 44.6 & 40.49 (f) & R & 419.6 \\
2E 1853.7+1534 & Sy 1 & 0.084 & 8.2 & GR & 1 & 44.25 & 44.56 & 38.86 & U & $<$ 39.27 \\
IGR J19378-0617 & NLS1 & 0.0106 & 6.8 & WL & 1 & 42.91 & 42.64 & 38.11 & U & $<$ 4.955 \\
NGC 6814 & Sy 1.5 & 0.0052 & 7.1 & WL & 1 & 40.98 & 42.52 & 37.07 & R & 27.88 \\
Cyg A & Sy 2 & 0.0561 & 9.4 & G & 1 & 44.68 & 44.73 & 40.69 (g) & R & 468.6 \\
4C 74.26 & Sy 1 & 0.104 & 9.6 & WL & 1 & 44.75 & 45.04 & 40.52 (h) & R & 1724.0 \\
S52116+81 & Sy 1 & 0.084 & 8.8 & WL & 1 & 44.25 & 44.77 & 40.71 & R & 701.6 \\
IGR J21247+5058 & Sy 1 & 0.02 & 6.6 & WL & 1 & 43.81 & 43.96 & 39.51 & R & 279.3 \\
SWIFT J2127.4+5654 & NLS1 & 0.014 & 7.2 & GR & 1 & 42.89 & 43.04 & 37.53 & S & 7.918 \\
RX J2135.9+4728 & Sy 1 & 0.025 & 7.5 & WL & 4 & 43.00 & 43.32 & 38.08 & U & $<$ 11.69 \\ 
NGC 7172 & Sy 2 & 0.0087 & 7.7 & WL & 1 & 42.51 & 43.11 & 37.87 & R & 36.88 \\
MR 2251-178 & Sy 1 & 0.064 & 6.9 & WL & 1 & 44.23 & 44.79 & 39.26 & U & $<$ 29.92 \\
MCG-02-58-22 & Sy 1.5 & 0.0469 & 7.1 & WL & 1 & 44.16 & 44.25 & 39.29 & U & $<$ 21.93 \\
IGR J23308+7120 & Sy 2 & 0.037 & 8.4 & $\mk$ & 2 & 42.80 & 43.55 & 37.92 & S & 40.31 \\
\hline
\end{tabular}
}
\begin{tablenotes}
\item {\it Notes:} Column 1 -- source name; Column 2 -- Seyfert classification; Column 3 -- redshift; Column 4 -- BH mass, Column 5 and 6 -- measurement method and reference for BH mass; Columns 7 and 8 -- X-ray luminosities in 2-10 keV ($\lx$) and 20-100 keV ($\lhx$); Column 9 -- radio luminosities at 1.4 GHz; Column 10 -- radio morphology; Column 11 -- length of half the major axis of the radio core. $\mbh$ in unit $\msun$, luminosities in unit $\ergs$, and $D_{\rm maj}$ in unit kpc.

\item {\it BH mass measurement method (Column 5).} RM: reverberation mapping; S: from stellar velocity dispersion; GR: from gas velocity and size of the broad-line region; WL: from the line width and luminosity of broad emission lines (e.g., H$\beta$, H$\alpha$, Pa$\beta$), sometimes the luminosity may be that at 5100\AA; G: from gas velocity dispersions (through e.g., [O {\tiny III}] or [Ne {\tiny III}]); $\mk$: from K-band magnitude/luminosity of either the whole host galaxy or the stellar bulge;

\item {\it Refences for BH mass (Column 6).} 1 -- \citetalias{Panessa2015}; 2 -- this work; 3 --\citet{Khorunzhev2012}; 4 -- The Swift/BAT AGN Spectroscopic Survey: \url{http://www.bass-survey.com/}.

\item {\it References for $\lr$ (Column 7).} (a) \citet{Kim2019}; (b) \citet{OrientiPrieto2010}; (c) \citet{Nagar2005}; (d) \citet{GirolettiPanessa2009}; (e) \citet{Middelberg2004}; (f) \citet{Dodson2008}; (g) \citet{StruveConway2010}; (h) \citet{Bourda2011}. The rest $\lr$ measurements are directly from \citetalias{Panessa2015}.

\item $^\star$ {\it The symbol of the radio morphology (see \citetalias{Panessa2015} for details).} U for unresolved, S for slightly resolved, R for resolved and A for ambiguous. Here, sources labelled with $^*$ are from the SUMSS survey, while the rest are from the NVSS survey.
\end{tablenotes}
\end{threeparttable}
\end{table*}

\section{Sample and Statistical Method} \label{sec:sample_stat}
\subsection{Sample}\label{sec:sample}

Our parent sample comes from \citetalias{Panessa2015}, who compiled a complete sample of moderately bright AGNs, selected at hard X-rays (20-100 keV) from the third {\it INTEGRAL}/IBIS survey. The sensitivity of {\it INTEGRAL}/IBIS above 20 keV is better than a few mCrab. Blazars are excluded in this sample, in order to avoid strong Doppler boosting effect (thus the luminosities measured are far away from their intrinsic values). The absorption-corrected 2-10 keV X-ray flux is from \citet{Malizia2009}, where they gather from literature. The radio data of the \citetalias{Panessa2015} sample mainly come from the NRAO VLA Sky Survey (NVSS; \citealt{Condon1998}) at 1.4 GHz, whose sensitivity is about $0.45$ mJy beam$^{-1}$. Additional complements are from the Sydney University Molonglo Sky Survey (SUMSS; \citealt{Bock1999}) at 843 MHz. In total there are 79 Seyfert galaxies in the \citetalias{Panessa2015} sample, among which 46 are Seyfert 1s (Seyfert 1-1.5, also include 6 narrow-line AGNs) and 33 are Seyfert 2s (Seyfert 1.9-2.0). 

For our investigation, detections/measurements of $\mbh, \lr$, and $\lx$ are necessary. We complement the $\mbh$ of some sources from other literature. For those that lack $\mbh$ estimation (noted as ``2'' in the column 6 of \autoref{tab1}), we follow \citet{Graham2007} to estimate their $\mbh$ as $\log(\mbh / \msun) = -0.37(\pm 0.04)\times(M_{\rm K} + 24) + 8.29(\pm 0.08)$. Here the absolute K-band magnitude of the galaxy $M_{\rm K}$ is obtained from Two-Micron All-Sky Survey (2MASS) through SIMBAD.\footnote{\url{http://simbad.u-strasbg.fr/simbad/}} This above empirical relationship has a total scatter of 0.33 dex in $\mbh$ \citep{Graham2007}, similar to those derived based on other methods. We exclude from our final sample sources that lack detections/measurements of either $\mbh, \lr$, or $\lx$. We further examined their radio fluxes at 1.4 GHz, to consider only the radio emission from the nuclei/core region. This is crucial for those resolved ones with extended structures. For those resolved ones, if possible, their nuclear radio emission at 1.4 GHz is then derived from higher-resolution (compared to VLA) VLBI\footnote{very long baseline interferometry, e.g., the {\it Very Long Baseline Array} (VLBA) and the {\it European VLBI Network} (EVN). } observations at frequencies specified below, assuming a radio spectrum $F_\nu \propto \nu^{-\alpha}$ with $\alpha = 0.7$ (the same as \citetalias{Panessa2015}), except for 2 sources that have reliable $\alpha$ measurements, i.e. NGC 1275 ($\alpha = -0.51$, \citealt{Kim2019}) and NGC 5506 ($\alpha = -0.06$, \citealt{Middelberg2004}). Compared to \citetalias{Panessa2015}, sources that have their radio fluxes updated are: NGC 1275 (based on VLBI observations at 43 GHz, \citealt{Kim2019}), NGC 3783 (1.6 GHz, \citealt{OrientiPrieto2010}), NGC 4151 (5 GHz, \citealt{Nagar2005}), NGC 4388 (1.6 GHz, \citealt{GirolettiPanessa2009}), NGC 5506 (1.6 GHz, \citealt{Middelberg2004}), 3C 390.3 (5 GHz, \citealt{Dodson2008}), Cyg A (1.34 GHz, \citealt{StruveConway2010}) and 4C 74.26 (2.3 GHz, \citealt{Bourda2011}). 

To summarize, of the original 79 AGNs in \citetalias{Panessa2015}, 9 (4) are excluded due to the lack of radio (BH mass) measurements. The other 2 are also excluded, who are only detected marginally or suffer background contamination in radio. Our final sample includes 64 sources, among which 35 are Seyfert 1s (Seyfert 1-1.5, 4 narrow-line Seyfert 1s are also included) and 29 are Seyfert 2s (Seyfert 1.9-2.0). We list in \autoref{tab1} the basic properties of our sample, including their Seyfert classification, redshift, black hole mass and the method adopted in its measurement, X-ray luminosities in 2-10 keV ($\lx$) and 20-100 keV ($\lhx$; from {\it INTEGRAL}/IBIS, see \citetalias{Panessa2015}), and radio luminosities at 1.4 GHz. Their radio morphology and the size of the radio core at 1.4 GHz (NVSS) or 0.8 GHz (SUMSS) are also provided for reference. 

Figure \ref{fig1} shows the basic properties of our sample. In both panels, the optical property is defined by the color and the shape, i.e. the Seyfert 1s are shown by green circles and the Seyfert 2s by orange triangles. All sources in our sample is fairly bright. The X-ray Eddington ratio $\lx/\ledd$ covers almost 4 orders of magnitude, i.e. between $\sim10^{-4.2}$ and $\sim10^{-0.3}$. For an X-ray bolometric correction $\lbol/\lx\approx 16$ (e.g., \citealt{Ho2008, Netzer2019}), the Eddington ratio ($\lambda_{\rm Edd} = \lbol/\ledd$) of our sample then ranges between $\sim10^{-3}$ and $\sim 10$. The left panel of Figure \ref{fig1} shows the BH mass distribution, where we find that $\mbh$ ranges between $10^{\sim5.5}\msun$ and $10^{\sim10}\msun$, with a clustering around $10^{7-9}\msun$. Moreover, there apparently exists a negative correlation between $\mbh$ and $\lx/\ledd$, where a simple power-law fit suggests that $\mbh \propto (\lx/\ledd)^{-1.33\pm 0.18}$. Pearson correlation coefficient of this fitting is listed in Table \ref{tab2}. The lack of small $\mbh$ AGNs at the low-$\lx/\ledd$ end is because our sample is limited by the X-ray flux; the lack of large $\mbh$ AGNs at the high-$\lx/\ledd$ end, on the other hand, is because of the cosmic evolution, i.e. galaxies in local universe, whose $\mbh$ is expected to be larger, are less active than those distant ones. 

We also separate radio-loud sources (open symbols in Figure \ref{fig1}) from radio-quiet ones (filled symbols), where the X-ray radio-loudness $R_{\rm X}$ is defined as $R_{\rm X}=\lr/\lx$ \citep{TerashimaWilson2003}\footnote{Note that conventionally the radio loudness $R$ is defined as the ratio of luminosities between 5 GHz radio band and the optical B-band (e.g., \citealt{Kellermann1989}).} and the RL/RQ boundary is set to  $R_{\rm X}=10^{-4.5}$. We emphasize that we do not observe,  in our near-complete hard X-ray selected AGN sample, the bimodal distribution of $R_{\rm X}$ (see also Figure 2 of \citetalias{Panessa2015}), suggesting that the bimodality in radio-loudness may be a selection effect.
Another property of our sample is that, as $\lx/\ledd$ increases, the fraction of RL sources declines significantly, i.e. there are only two RL AGNs in the $\lx/\ledd > 10^{-2}$ regime. One is IGR J10404-4625 whose $\lx/\ledd\approx 0.02$, and the other is IGR J21247+5058 whose $\lx/\ledd\approx 0.13$.

\subsection{Statistical Method and Basic Assumptions}\label{sec:method}

As clearly shown in the left panel of Figure\ \ref{fig1}, there is a strong $\mbh-\lx/\ledd$ relationship, which obviously contaminates the estimation of $\xim$ in Equation \ref{eq_fp}. To avoid this technical problem, we in this work aggressively omit the dependence on $\xim$ but focus on $\xix$, i.e. we consider the following radio/X-ray correlation in Eddington unit,
\begin{equation}
\log(\lr/\ledd) = \xix \log(\lx/\ledd) + const.. \label{eq_rx}
\end{equation}
Since most sources have a clustering of $\mbh$ at $\sim 10^{7-9}\msun$, physically it is equivalent to the case of absorbing the impact of $\mbh$ into $const.$.

Following \citet{Merloni2003}, we statistically fit the observational data through the minimization of the following quantity (hereafter the least $\chi^2$ approach),
\begin{equation}
\chi^2 = \Sigma {\left(\log(\lr/\ledd) - \xix \log(\lx/\ledd) - const.\right)^2\over (\sigma_{\rm R,Edd} \log(\lr/\ledd))^2 + (\xi_{\rm X} \sigma_{\rm X,Edd} \log(\lx/\ledd))^2}. \label{chi2}
\end{equation}
Considering the non-simultaneity of the radio and X-ray fluxes used in this work, we  ignore the observational uncertainties in $\log(\lr/\ledd)$ and $\log(\lx/\ledd)$, but directly take their uncertainties to be $\sigma_{\rm R,Edd} = 0.3$ and $\sigma_{\rm X,Edd}=0.3$, respectively (see e.g., \citealt*{Merloni2003, Gultekin2009, XieYuan2017}).

We note that in Equation \ref{chi2} we additionally weight the uncertainties by luminosities in Eddington unit (i.e., $\lr/\ledd, \lx/\ledd$). This is equivalent to an emphasis on fainter sources. We test this revised regression method through data sets which are manually generated by given parameters of $\xix$, $\sigma_{\rm R,Edd}$ and $\sigma_{\rm R,Edd}$. We find that the regression based on Equation \ref{chi2} provides a better recovery of the input $\xix$ value than that by original formulae of \citet{Merloni2003}.

\begin{figure}
  \includegraphics[width=8cm]{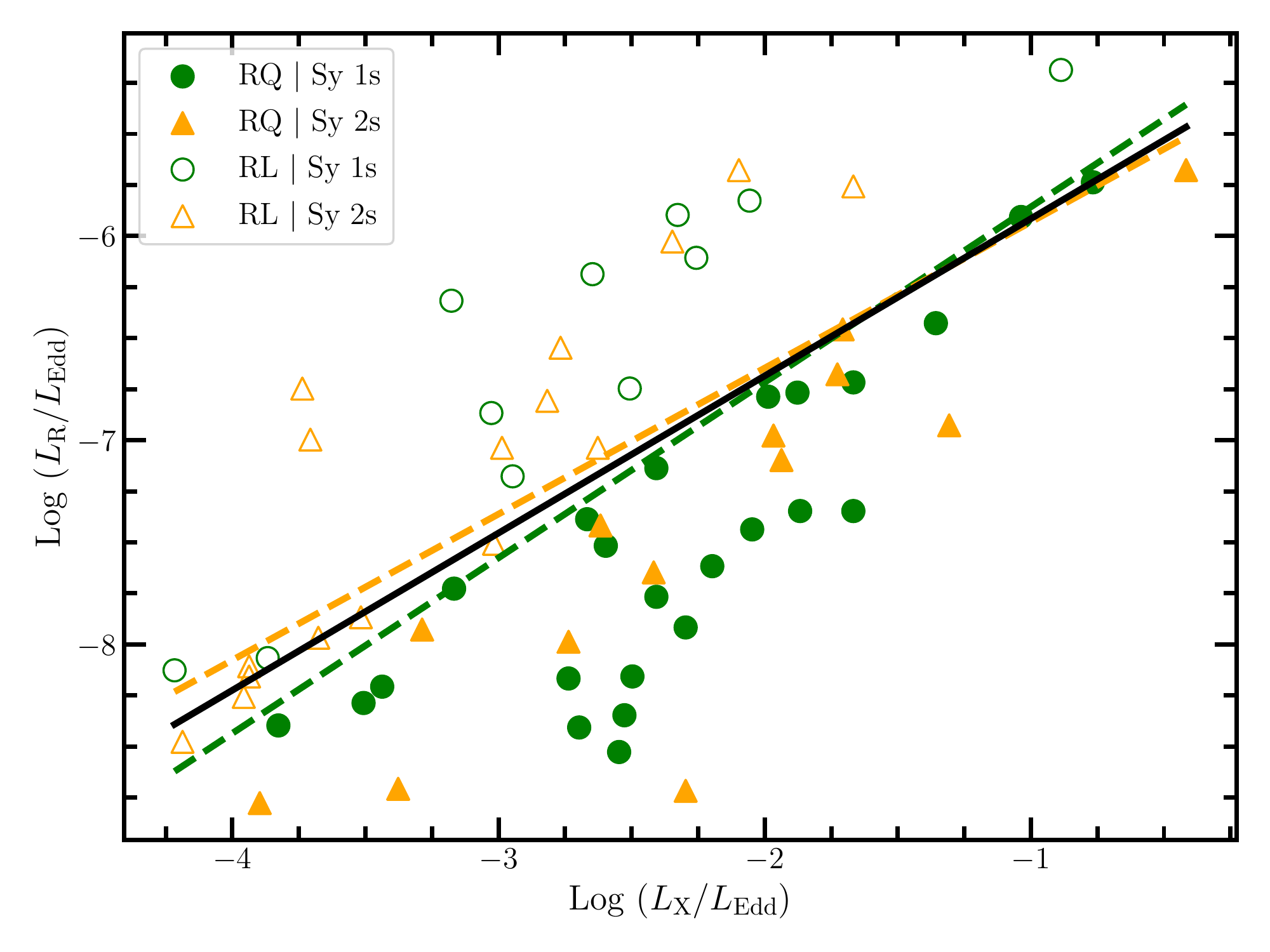}
  \caption{Radio/X-ray correlation (in Eddington units) for Seyfert galaxies, with a comparison between Seyfert 1s (the green filled circles) and Seyfert 2s (the orange filled triangles). The black solid curve represents the fitting to the whole sample (see Equation \ref{rx_all}), while the dashed curves provide the fitting results of two subsamples (see Equation \ref{sy12}), with the color the same to those of the data points.}
  \label{fig2}
\end{figure}

\section{Results}\label{sec:result}

\subsection{X-ray flux ratio $\lhx/\lx$ distribution}

We first investigate the distribution of X-ray flux ratio $\lhx/\lx$ as a function of $\lx/\ledd$. We caution that $\lhx$ and $\lx$ are not observed in a coordinated manner, the time interval can be as large as several years (\citealt*{Malizia2009}; \citetalias{Panessa2015}). AGNs are known to be variable on year timescale \citep{UlrichMaraschiUrry1997, Netzer2008}, thus $\lhx/\lx$ may suffer this non-simultaneity issue. 

As shown in the right panel of Figure \ref{fig1}, we find that most sources have $\lhx/\lx\sim 1-4$, despite the non-simultaneity issue. Moreover, statistically there is no dependence of the ratio $\lhx/\lx$ on the X-ray luminosity $\lx/\ledd$. Furthermore, we separate the sources according to their radio-loudness and Seyfert classification. Still, we do not observe any separation in $\lhx/\lx$ between RL and RQ ones, nor between Seyfert 1s and Seyfert 2s. All these results may imply that, despite the four-order-of-magnitude dynamical range in $\lx/\ledd$, the possible change in accretion mode (i.e. cold accretion at $\lx/\ledd \ga 10^{-3}$ versus hot accretion at $\lx/\ledd \la 10^{-3}$), and the huge difference in their radio loudness and/or Seyfert classification, all the sources in our sample have the same radiative mechanism and origin in X-rays.\footnote{We note that, the jet may also dominate the X-rays in RL AGNs \citep{HarrisKrawczynski2006, BlandfordMeierReadhead2019}; and radio origin may also be diverse \citep{Panessa2019, chiaraluce2020}. See also {\it Introduction}.} From a theoretical point of view, the most plausible mechanism is hot accretion flow (or corona in case of cold accretion).

\subsection{Radio/X-ray Correlation in Seyfert Galaxies} \label{sec:rx}

The key motivation of this work is to investigate $\xix$, i.e. the radio/X-ray correlation slope, among different types of Seyfert AGNs. We take three different classification methods, i.e. according to, 1) the full width at half-maximum (FWHM) of the emission lines in optical band, 2) the radio loudness $R_{\rm X}$, and 3) the size of radio-emission site. 

\subsubsection{impact of optical emission-line properties}
We first consider their differences in optical emission-line properties, i.e. Seyfert 1s versus Seyfert 2s. There are 35 (${\sim}52\%$) Seyfert 1s  and 29 (${\sim}48\%$) Seyfert 2s. We plot in the Figure\ \ref{fig2} all the sources in the revised ($\log(\lr/\ledd), \log(\lx/\ledd)$) plane, where Seyfert 1s are shown by green filled circles and Seyfert 2s are shown by orange filled triangles. 

Several results can be derived immediately. First, the observations do exhibit large scatters (see also \citealt*{Merloni2003,Falcke2004}), i.e. at a given X-ray luminosity $\lx/\ledd$, the nuclear radio luminosity $\lr/\ledd$ can differ by 2-3 orders of magnitude, and the scatter is similar among Seyfert 1s and Seyfert 2s. This may suggest as a piece of evidence against the AGN unification model, where viewing angle is considered as the primary factor, i.e. face-on for Seyfert 1s and edge-on for Seyfert 2s  (see, e.g., \citealt*{Netzer2015, Padovani2017} for a summary of additional evidence against the unification model).\footnote{Indeed some individual AGNs, so-called changing-look AGNs, can change their appearances in optical emission lines (i.e. between type 1 with broad lines and type 2 with only narrow lines) over several years (e.g., \citealt{LaMassa2015}), a period that the orientation should not vary much.} Actually, we notice from this plot that Seyfert 1s and Seyfert 2s overlap each other in $\lr/\ledd$, suggesting that beaming effect is not the key factor. Based on our current understanding of jet physics, we may argue that additional factor(s), among which likely the magnetic flux (\citealt{Sikora2007}) and/or the BH spin (\citealt{UnalLoeb2020}), should play dominate roles in introducing the scatters as observed.

Another result from this plot is that, the Seyfert 1s and Seyfert 2s cover a similar range in both radio and X-ray luminosities (in Eddington unit), although as shown in Figure\ \ref{fig1} the Seyfert 2s are less massive in $\mbh$ compared to Seyfert 1s, especially at the bright $\lx/\ledd$ regime. 

We fit the data under the least $\chi^2$ approach. As shown by the black solid curve in Figure\ \ref{fig2}, the whole sample follows
\begin{equation}
\log(\lr/\ledd) = 0.77^{+0.10}_{-0.10} \log(\lx/\ledd) - 5.15^{+0.25}_{-0.25}.\label{rx_all}
\end{equation}
Meanwhile, the Seyfert 1s and Seyfert 2s follow, respectively,
\begin{eqnarray}
\log(\lr/\ledd) & = & 0.86^{+0.15}_{-0.15} \log(\lx/\ledd) - 5.00^{+0.36}_{-0.36}, \hspace{0.3 cm} {\rm (Sy\,1s)} \nonumber\\
\log(\lr/\ledd) & = & 0.72^{+0.13}_{-0.13} \log(\lx/\ledd) - 5.22^{+0.35}_{-0.35}.\hspace{0.3 cm} {\rm (Sy\,2s)} \nonumber\\
& &\label{sy12}
\end{eqnarray}
The coefficients of Pearson correlation analysis are listed in Table \ref{tab2}. The slopes in the two subsamples agree with each other within $\sim$1$\sigma$. This implies that, either orientation is not the key difference between Seyfert 1s and Seyfert 2s (thus disfavour the AGN unification model), or there is no intrinsic difference in the origin of X-ray emission.

\subsubsection{impact of radio-loudness}

We then explore the impact of radio-loudness, which is widely considered as an indicator of jet beaming effect. According to our RL/RQ separation ($R_{\rm X}=10^{-4.5}$; \citealt*{TerashimaWilson2003}), 27 out of 64 (${\sim}42\%$) sources are RL and the rest 37 out of 64 (${\sim}58\%$) sources are RQ. Such a high fraction of RL sources (compared to the typical ${\rm}10\%$ in bright AGNs and quasars) is known in literature (e.g., \citealt*{Ho2008}). Because of the classification, RL AGNs are systematically brighter in radio luminosity $\lr/\ledd$, at a given X-ray luminosity $\lx/\ledd$. 

\begin{figure}
  \includegraphics[width=8cm]{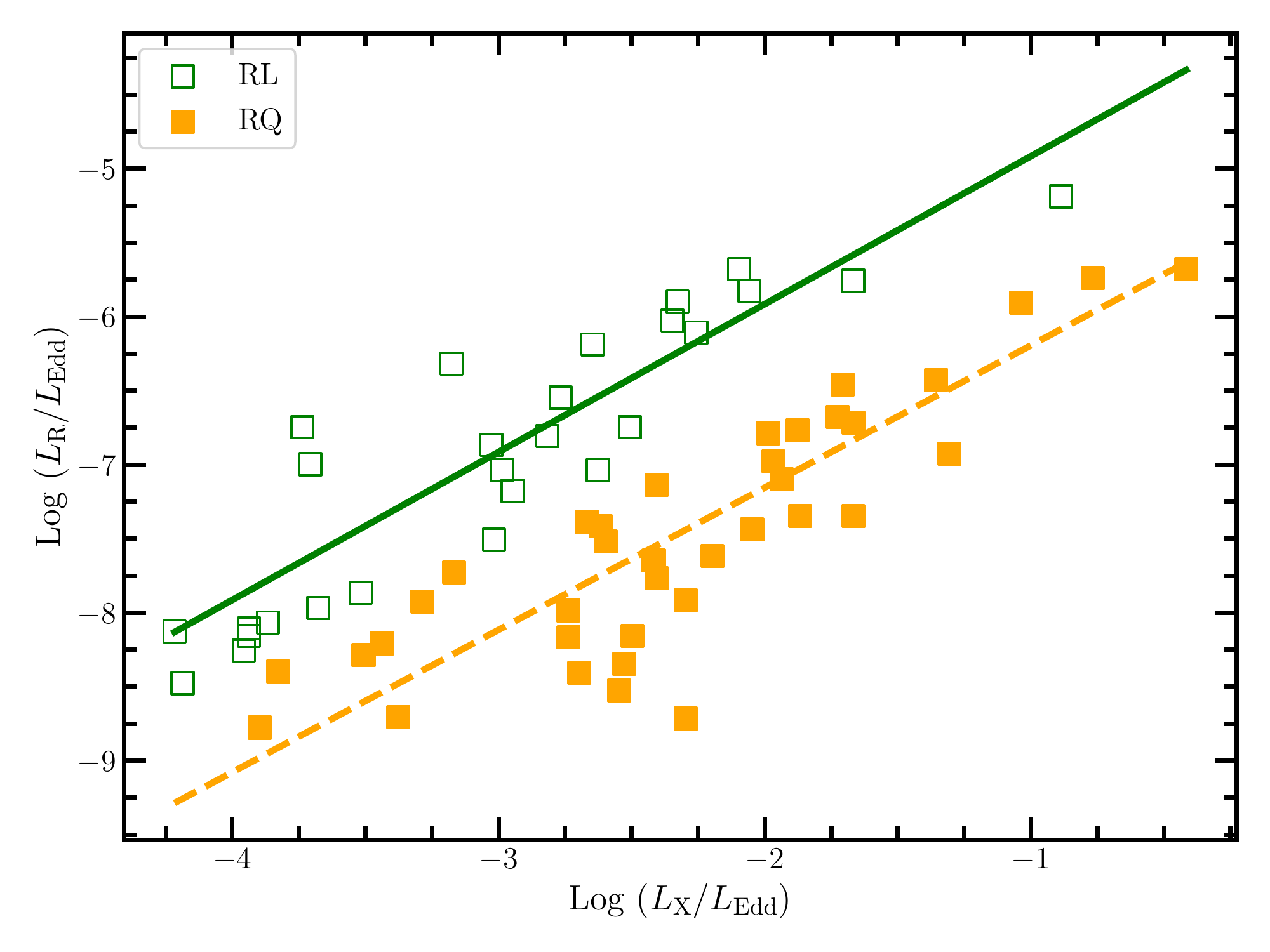}
  \caption{Radio/X-ray correlation (in Eddington units) for Seyfert galaxies with different radio loudness $R_{\rm X}$. Here the green open squares are for RL AGNs and the orange filled squares for RQ ones. The fitting results are also shown (see Equation \ref{eq:rlrq}), with the color the same to that of the data points.}
  \label{fig3}
\end{figure}

We re-plot the sample in Figure \ref{fig3}, where RL AGNs are shown by green open squares and RQ AGNs are shown by orange filled squares. We find that most RL AGNs in our sample are less luminous in X-rays,  only two out of 27 RL AGNs (${\sim}7\%$), i.e. IGR J10404-4625 and IGR J21247+5058, is bright with $\lx/\ledd > 10^{-2}$. On the other hand, for the 14 bright RQ AGNs with $\lx/\ledd > 10^{-2}$, there is no preference in Seyfert types (see the left panel of Figure \ref{fig1}), i.e. 8 sources are Seyfert 1s and the rest 6 sources are Seyfert 2s.

We fit the data with least $\chi^2$ regression method and find that RL and RQ AGNs follow, respectively,
\begin{eqnarray}
\log(\lr/\ledd) & = & 1.00^{+0.09}_{-0.09} \log(\lx/\ledd) - 3.92^{+0.25}_{-0.25}, \hspace{0.3 cm} {\rm (RL\,AGNs)} \nonumber\\
\log(\lr/\ledd) & = & 0.96^{+0.08}_{-0.08} \log(\lx/\ledd) - 5.23^{+0.18}_{-0.18}. \hspace{0.3 cm} {\rm (RQ\,AGNs)}\nonumber\\
& & \label{eq:rlrq}
\end{eqnarray}
The coefficients of Pearson correlation analysis are listed in Table \ref{tab2}. Again, the correlation slopes of RL and RQ AGNs are consistent with each other at 1$\sigma$ uncertainty level, 

Our result of a $\xix\sim 1$ FP in RL AGNs agrees with that reported in \citet{Liao2020}. They investigated a sample of young radio AGNs, which all have radiatively efficient accretion (e.g., cold accretion) and powerful radio emission.
Most previous work find that RL AGNs follow a much steeper relationship, i.e. $\xix\sim 1.3$ (e.g., \citealt*{Panessa2007, LiWuWang2008, deGasperin2011}), and the correlation becomes steeper (i.e., larger $\xix$) as the radio-loudness increases \citep{LiWuWang2008}. In accretion theory, the standard FP is achieved when the hot accretion flow is responsible for the X-rays \citep{YuanCui2005, XieYuan2016}; the steep $\xix\sim 1.3$ FP can be observed when the X-rays originates from the relativistic jet rather than the hot accretion flow (see e.g., discussions in \citealt{YuanCui2005, XieYuan2017}). Such a situation can only be achieved under two conditions, one is when the system is faint enough (e.g., $\lx/\ledd\lesssim 10^{-6}$; \citealt*{YuanCuiNarayan2005,XieYuan2017}), and the other is when the beaming effect is sufficiently strong (high Doppler boosting effect for the jet emission in X-rays; e.g., \citealt*{Panessa2007, LiWuWang2008}). In this theoretical picture, our finding of a universal $\xix$ among RL and RQ systems suggests that the X-rays of RL Seyferts also originate from hot accretion flow, the same as RQ Seyferts. This interpretation is also favoured by the $\lhx/\lx$ distribution, as shown in Figure\ \ref{fig1}. We note that similar results have been reported in literature. For example, based on a compilation of 13 low-excitation radio galaxies (LERGs; belong to RL AGNs in our classification), \citet{LiGu2018} find the LERGs to follow the standard $\xix$ correlation, not a steep one as reported in \citet*{deGasperin2011}. The difference is that in the latter work they also include LERGs which have a steep radio spectrum.

Finally, we caution that the RQ systems here also agree with the linear $\lr/\lx\sim10^{-5}$ relationship observed in corona-active stars, thus being explained under an AGN corona (above the SSD) model (\citealt*{LaorBehar2008}; see Sec\ \ref{sec:rqmodel}).

\begin{figure}
  \includegraphics[width=8cm]{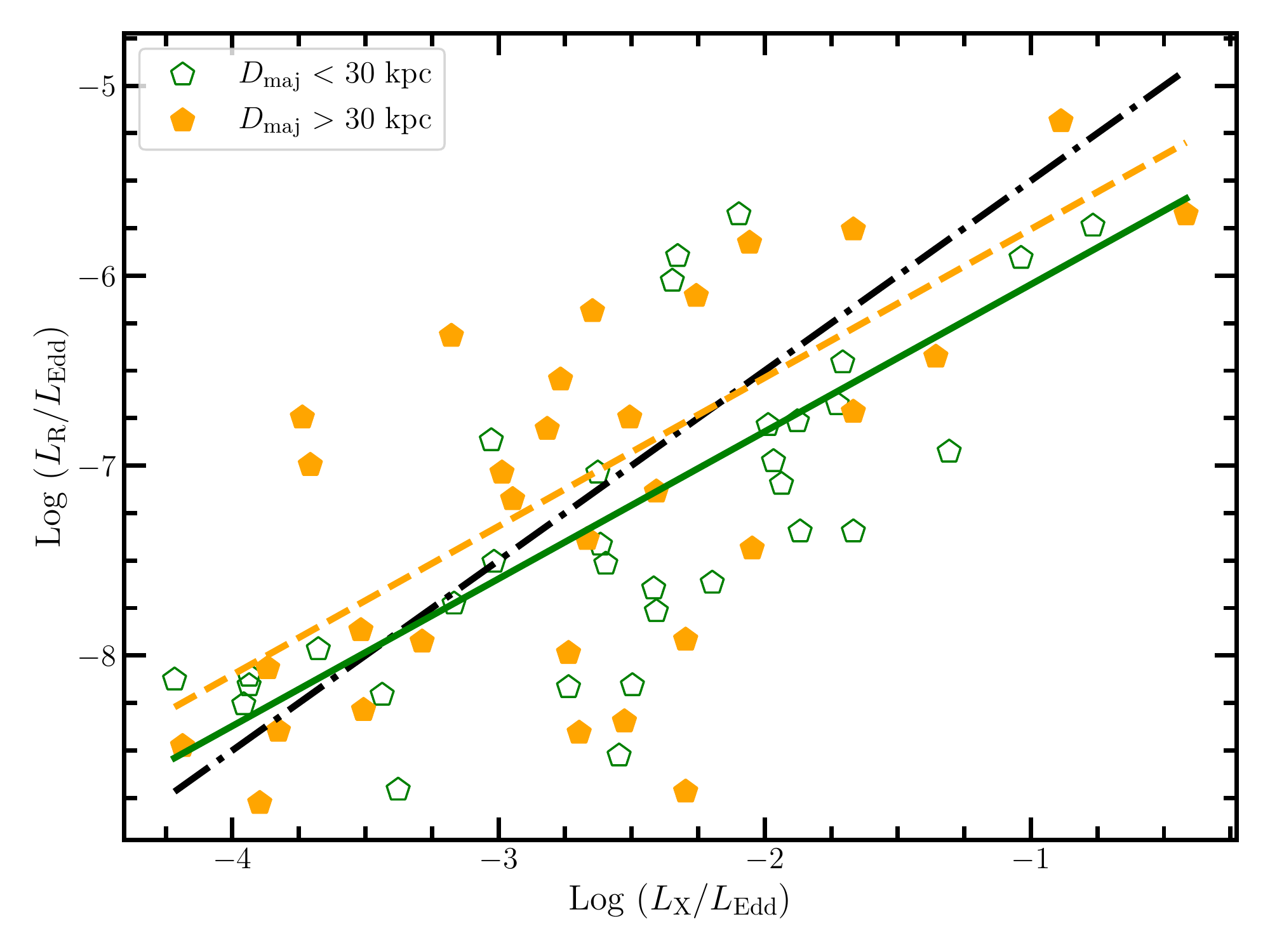}
  \caption{Radio/X-ray correlation (in Eddington units) for Seyfert galaxies with different radio emission size. The green open pentagons and orange filled pentagons are respectively, for sources with their radio size (length of the half major axis) are less or greater than 30 kpc. The fitting results are also shown, with the color the same to those of the data points.}
  \label{fig4}
\end{figure}

\subsubsection{impact of emission size in radio}
We finally investigate the impact of the radio emission size. It is naturally expected that, the velocity and power of jet, which can be illustrated by the radio morphology, are crucial to determining the interaction between the jet and the host galaxy (e.g., \citealt*{McNamaraNulsenN2012, DuanGuo2020}). Again we caution that the jet interpretation of RQ AGNs may be an over-simplification, see Sec.\ \ref{sec:rqmodel}. More than 50\% sources in our sample do not show clear extended structure.\footnote{It is well-known that jet in RL AGNs morphologically has a Fanaroff–Riley dichotomy distribution, with the core-dominate FR I jets being shorter and less stable than the lobe-dominate FR II jets \citep{FanaroffRiley1974}.} The only source with FR I jet morphology is NGC~1275, and the 5 sources with FR II jet morphology are respectively, B3 0309+411, 3C 111, 3C 390.3, Cyg A and 4C 74.26. We thus take the length of half the major axis $D_{\rm maj}$ as a representative of the radio morphology. We take observations of either NVSS (at 1.4 GHz) or SUMSS (at 0.8 GHz) (see Sec.\ \ref{sec:sample} and Table\ \ref{tab1}), and define jets whose $D_{\rm maj}<30$ kpc as compact systems, and those whose $D_{\rm maj}>30$ kpc as extended (or resolved) systems. Under this definition, 31 Seyferts in our sample belong to the extended category, and the rest 33 Seyferts belongs to the compact category.

We show the results in Figure\ \ref{fig4}, where the green open pentagons and orange filled pentagons are respectively, for sources with their $D_{\rm maj}$ are less or greater than 30 kpc. Obviously there is no clear separation in the $\lr/\ledd-\lx/\ledd$ plane between systems with these two types of radio morphologies, i.e. they overlap with each other. Moreover, with a comparison to Figure\ \ref{fig3}, we find that there is no direct connection between radio emission size and the radio-loudness, i.e. 11 out of 27 ($41\%$) RL AGNs remain compact with $D_{\rm maj}<30$ kpc; and 15 out of 37 ($41\%$) RQ AGNs show extended radio emission. When $\lx/\ledd\ga 10^{-2}$, there is a weak tendency that Seyferts with extended radio morphology are more luminous in $\lr/\ledd$ (equivalently, radio-louder) at a given $\lx/\ledd$. But the sample size in this region is small, and such tendency disappears when $\lx/\ledd\la 10^{-2}$.
We further fit the data with least $\chi^2$ regression method and find that Seyferts with compact and extended radio emission follow respectively (see Table \ref{tab2} for Pearson correlation coefficients),
\begin{eqnarray}
\log(\lr/\ledd) & = & 0.78^{+0.13}_{-0.13} \log(\lx/\ledd) - 5.27^{+0.32}_{-0.32}, \hspace{0.3 cm} {\rm (Compact)} \nonumber\\
\log(\lr/\ledd) & = & 0.78^{+0.14}_{-0.14} \log(\lx/\ledd) - 4.97^{+0.37}_{-0.37}. \hspace{0.3 cm} {\rm (Extended)}\nonumber\\
& &
\end{eqnarray}
Clearly, the correlation slopes are consistent with each other within 1$\sigma$ uncertainty level.

Finally, we note that we also check different choices of $D_{\rm maj}$ criteria, i.e. 20 kpc or 25 kpc, and the results (not shown here) are quite similar to that reported above.

\begin{table}
	\caption{Pearson correlation analysis of each subsample}
	\centering
	\label{tab2}
	\begin{threeparttable}
		\begin{tabular}{rccc}
			\hline
			Sample (No.)		& Pearson	& $P_{\rm null}$		& $R^2$	\\
			\hline
			\noalign{\smallskip}
			&\multicolumn{3}{c}{Figure \ref{fig1}: log ($\mbh$) vs. log ($\lx/\ledd$)}\\
			\noalign{\smallskip}
			full sample (64) & $-0.54$	& $3.2 \times 10^{-6}$	& 0.54\\
			\noalign{\smallskip}
			&\multicolumn{3}{c}{Figure \ref{fig2}--\ref{fig4}: log ($\lr/\ledd$) vs. log ($\lx/\ledd$)}\\
			\noalign{\smallskip}
			full sample (64) 	& $0.66$	& $2.9 \times 10^{-9}$	& 0.49\\
			Sy 1s (35)			& $0.65$	& $2.6 \times 10^{-5}$	& 0.56\\
			Sy 2s (29) 			& $0.67$	& $3.9 \times 10^{-5}$	& 0.45\\
			RQ (37) 			& $0.86$	& $9.2 \times 10^{-12}$	& 0.76\\
			RL (27) 			& $0.91$	& $7.0 \times 10^{-11}$	& 0.81\\
			Compact (33) 		& $0.71$	& $3.3 \times 10^{-6}$	& 0.56\\
			Extended	(31) 	& $0.64$	& $1.2 \times 10^{-4}$	& 0.45\\
			\noalign{\smallskip}
			&\multicolumn{3}{c}{Figure \ref{fig5}: log ($\njet$) vs. log ($\lx/\ledd$)}\\
			\noalign{\smallskip}
			full sample (64) 	& $-0.44$	& $2.9 \times 10^{-4}$	& 0.36\\
		\hline
	\end{tabular}
	\begin{tablenotes}
	\item {{\it Notes:} Column 1 -- sample; Column 2 and 3 -- Pearson correlation coefficient and null correlation hypothesis probability; Column4 -- goodness of fit, defined as ESS (Explained Sum of Squares) divided by TSS (Total Sum of Squares). The closer the $R^2$ is to 1, the better the fitting is. Check \url{http://online.sfsu.edu/mbar/ECON312_files/R-squared.html} for detail definition of the $R^2$. }
	\end{tablenotes}
	\end{threeparttable}
\end{table} 

\section{Discussions}\label{sec:discuss} 


\subsection{Alternative Models for the Radio and X-ray emissions in RQ AGNs}\label{sec:rqmodel}

One notable limitation of this work is that, we attribute in this work (a dominant fraction of) the nuclear radio emission to be from the jet component, either well-collimated or weakly-collimated. Due to the low resolution of NVSS and large $D_{\rm maj}$ in our sample, alternative contributions indeed cannot be ruled out, especially in RQ systems.

Unlike RL systems, many RQ Seyferts do not show jet-like structure in high-resolution radio observations (e.g., \citealt*{Ulvestad2005} and references therein): some remain unresolved at sub-pc scale, while others are extended but lack linear (jet-like) morphology. We notice that the origin of nuclear radio and/or X-ray emission in RQ Seyferts is actually under active debate in recent years (e.g., \citealt*{Bonchi2013, Baek2019, Laor2019, Smith2020, Fischer2021}, and references therein.); alternative origins of radio emission in RQ AGNs include star formation, AGN wind and AGN corona (see e.g., \citealt*{Padovani2017, Panessa2019} for reviews). Below we provide brief discussions on these alternative origins. 

One origin is the nuclear star formation activities. Active star formation processes will provide extended emission in radio and near-infrared. Based on a VLA 22 GHz RQ AGN sample, \citet{Smith2020} report that, after core flux subtraction, AGN with compact morphology will drop below the star formation expectation. This implies that the radio emission from nuclear star formation is still unresolved even at 1 arcsecond resolution. Higher resolution (or at higher frequency) observations are thus necessary. However, AGN activities co-evolve with the dynamics of the circum-nuclear medium, where a tight connection to nuclear star formation is a direct signature (e.g., \citealt*{Zhuang2021}). In this understanding, the star formation scenario for the non-linear radio emission awaits additional evidence. We also emphasize that the dominant mechanism for radio emission at different frequencies might be different (e.g.,  in star-forming galaxies, free--free emission will become comparable important to synchrotron at high frequency).

Another scenario proposed in literature is a hot corona (above the cold SSD) model for not only X-rays but also radio \citep{LaorBehar2008} in the RQ systems. This model is similar to the corona in stellar systems, where a linear $\lr/\lx \sim 10^{-5}$ relationship is established  \citep{GuedelBenz1993}. Observationally the RQ sources seem to be broadly consistent with this model (e.g., \citealt{Smith2020} and references therein). We note that our interpretation shares the same origin of X-rays, but differs in the origin of radio emission, i.e. in their model from non-thermal or hot thermal electrons in the corona \citep{LaorBehar2008, RaginskiLaor2016}, or in our interpretation from non-thermal electrons in the weakly-collimated jet. One supporting evidence of wind or corona scenario is from \citet{LiWuWang2008}, where they found that the broad line luminosity is tight correlated with radio luminosity in RQs, i.e. the radio emission in RQs might be nearly isotropic.

Our result suggests a universal radio/X-ray correlation in Seyferts, irrelevant to radio-loudness. However, the uncertainties in $\xix$ are still large, thus although we favor the jet interpretation, we actually cannot rule out any alternative scenarios, especially because all these processes correlate with each other. In principle, the contribution of nuclear star formation process can be removed from regular monitoring of individual sources, since the AGN and stellar processes have different variability timescale; the corona and wind/outflow models, on the other hand, are tightly correlated to the accretion process, thus challenge to discriminate.

\subsection{Jet Production Efficiency $\eta_{\rm jet}$}\label{sec:eta_jet}

One important quantity in accretion theory is the jet production efficiency $\eta_{\rm jet}$, which characterizes the fraction of accretion power that enters into the relativistic jet (e.g., \citealt{vanVelzenFalcke2013, Ghisellini2014, Inoue2017, Rusinek2020, SoaresNemmen2020, Wojtowicz2020}). It is defined as,
\begin{equation}
\eta_{\rm jet} = {P_{\rm jet}\over \dot{M} c^2}. \label{eta_jet}
\end{equation}
Obviously, the jet production efficiency $\eta_{\rm jet}$ will also dependent on magnetic flux attached to jet and the BH spin. Based on a sample of $\sim 7000$ radio-loud quasars, \citet{Inoue2017} reported an efficiency of $\eta_{\rm jet}\approx 1\times10^{-2}$ in the cold accretion disc regime, suggesting the BH spin and/or the magnetic flux are low in these systems; while based on a sample of $\sim 200$ well-selected blazars with $\gamma$-ray detection, \citet*{Ghisellini2014} find that their $\njet\sim 1-10$.

Here we investigate the jet production efficiency of Seyfert galaxies. There are several methods to estimate the jet power (see \citealt*{SoaresNemmen2020} for a brief summary), i.e. the radio lobe emission based on equipartition assumption \citep{Willott1999}, the radio core-shift effect \citep{ShabalaSantosoGodfrey2012} and the spectrum modelling \citep{Ghisellini2014}. In this work we follow that of \citet*{Willott1999} (for later updates, see e.g., \citealt*{Cavagnolo2010}). Based on a sample of 77 RL AGNs that includes both FR Is and IIs, they found that the jet power has a strong correlation with the radio luminosity at 151 MHz, which can be re-expressed as $\pjet \approx 1.9\times10^7\, f^{3/2}\, \lr^{6/7}\,{\ergs}\approx 6.0\times10^8 \lr^{6/7}\,{\ergs}$ at 1.4 GHz, if a radio spectrum $F_\nu \propto \nu^{-0.7}$ is adopted. The theoretical uncertainties in $\pjet$ are absorbed in a parameter $f$, which we take $f = 10$, following the recent calibrations by studies of X-ray cavity and hot spot in AGNs \citep{GodfreyShabala2013}. We caution that the calibration implies that the derived jet power is a time-averaged (over the past millions of years) one, and may differ from the $\pjet$ of the current on-going accretion process (represented by $\lx$). We for simplicity omit this uncertainty. We further caution that the above jet power estimation is from RL AGNs; the application to RQ AGNs may be aggressive and risky, with uncertainties hard to measure. However, this is the only method we currently have.

The accretion power $\dot{M} c^2$ can be estimated from the bolometric luminosity of hot accretion flow $L_{\rm bol,h}$\footnote{Expressed as the bolometric disc luminosity, i.e. $L_{\rm disc}$, in the notation of \citet*{Inoue2017}.} as $\dot{M} c^2=L_{\rm bol,h}/\epsilon$, where $\epsilon$ is the radiative efficiency. For a geometrically-thin cold accretion disc like SSD \citep{ShakuraSunyaev1973}, the efficiency is $\epsilon_{\rm SSD}\approx 10\%$; while for a hot accretion flow, the efficiency is systematically lower than $\epsilon_{\rm SSD}$, but its value increases as $\dot{M}$ increases (\citealt*{XieYuan2012, XieZdziarski2019}), and can be comparable to that of cold SSD at high $\dot{M}/\dot{M}_{\rm Edd}$ end. We further estimate the $L_{\rm bol,h}$ from the X-ray bolometric correction factor $f_{\rm X} = L_{\rm bol,h}/\lx$ (e.g., \citealt*{Ho2008,VasudevanFabian2009}). Then we can re-express Equation (\ref{eta_jet}) as,
\begin{equation}
{\njet\fx\over\epsilon} = {P_{\rm jet}\over\lx}\propto {L_{\rm R}^{6/7}\over\lx}. \label{eq:eta_jet_combine}
\end{equation}

\begin{figure}
\vspace{-0.3cm}
\includegraphics[width=8cm]{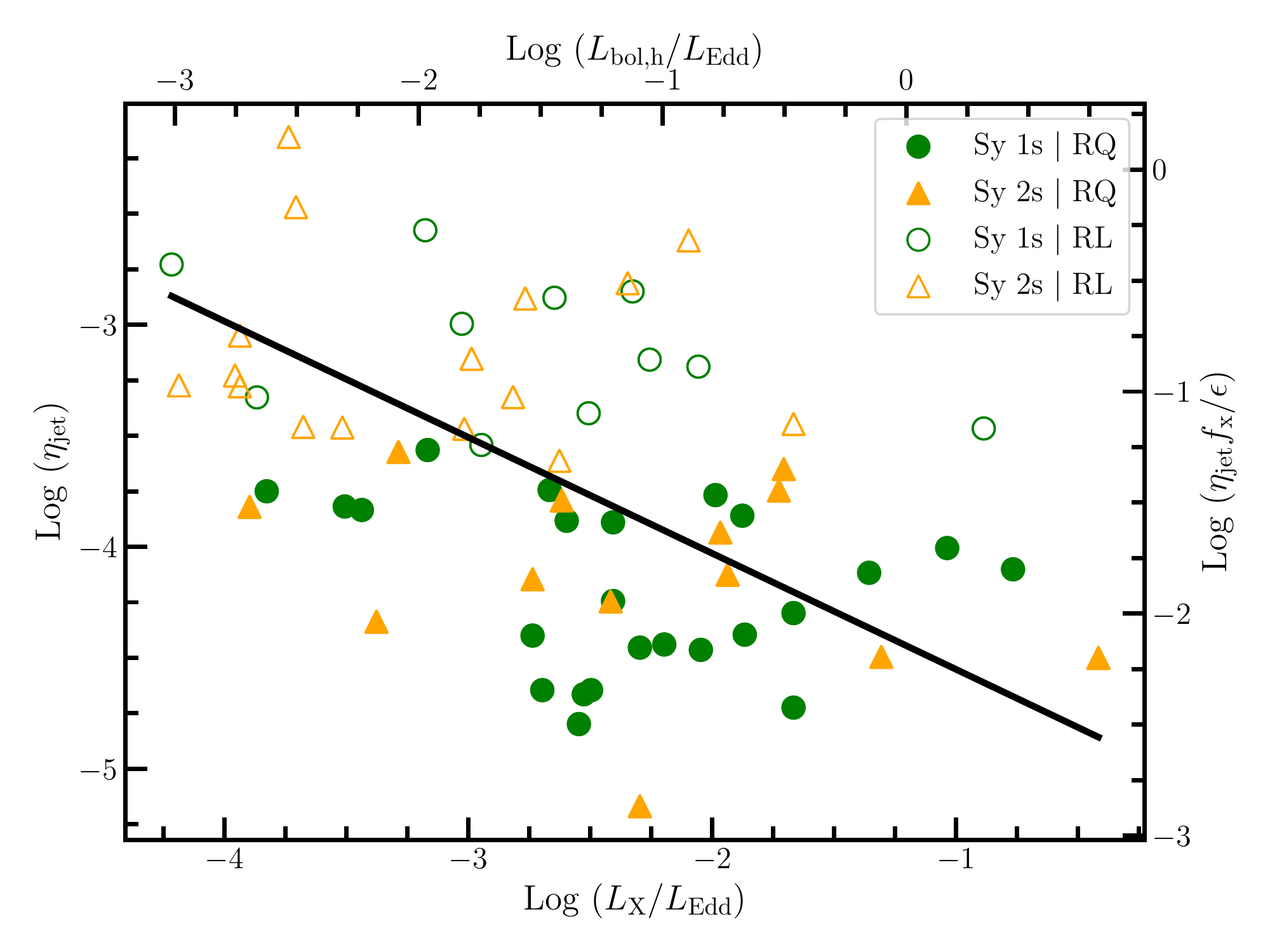}
\caption{Jet production efficiency $\njet$ as a function of the X-ray Eddington ratio $\lx/\ledd$ (and in the upper x-axis the bolometric Eddington ratio $L_{\rm bol,h}/\ledd$), where we assume the radiative efficiency of hot accretion flow $\epsilon=8\%$ and the X-ray bolometric correction factor $\fx=16$. The right y-axis shows $\njet \fx / \epsilon\equiv P_{\rm jet}/\lx$. The black solid curve represents the fitting result to the whole sample, see Equation (\ref{eq:eta_lbol}). As labelled in the plot, the symbols are of the same meaning to those in Figure \ref{fig1}. }
\label{fig5}
\end{figure}


We plot in Figure \ref{fig5} the relationship between $\lx/\ledd$ and the jet production efficiency $\njet$, where the symbols are of the same meaning to those in Figure \ref{fig1}. Here we statistically assume  $\fx \simeq 16$ \citep{Ho2009} and $\epsilon \simeq 8\%$ \citep{XieYuan2012}. The original value of $\njet \fx/\epsilon$ ($\equiv P_{\rm jet}/\lx$, see Equation \ref{eq:eta_jet_combine}) of each source is also shown in Figure \ref{fig5} by its right y-axis. Several results can be derived directly from this plot. First, $\njet$ (or more accurately, $\njet \fx / \epsilon$) shows a pretty large scatter at given X-ray luminosity ($\lx/\ledd$), and the scatter is similar among RL AGNs and RQ AGNs. On the other hand, there is a tendency that Seyfert 1s has a smaller scatter than Seyfert 2s. For the whole sample of 64 Seyferts, we find that $\njet$ varies between $10^{-5.2}$ and $10^{-2}$, with a mean value of ${<}\njet{>}=1.9^{+0.9}_{-1.5} \times 10^{-4}$. We note that compared to the results in literature  \citep{vanVelzenFalcke2013, Ghisellini2014, Inoue2017, SoaresNemmen2020, Wojtowicz2020}, the scatter is roughly consistent, but the mean value derived here is at least two orders of magnitude smaller. The mean jet production efficiency for RL and RQ subsamples are respectively, ${<}\njet{>}=7.9^{+1.9}_{-4.5} \times 10^{-4}$ and ${<}\njet{>}=6.9^{+1.7}_{-4.1} \times 10^{-5}$. We note that, based on a large sample of low-redshift AGNs selected from the Swift/BAT, \citet*{Rusinek2020} recently find a mean production efficiency of ${<}\njet{>}\approx2\times10^{-5}\, (\epsilon/10\%)$. Part of the reason for such low ${<}\njet{>}$ is because that, about $80\%$ of the sources in their sample are RQ AGNs, which by definition are systems with low $\njet$.

Second, even within this limited (but near-complete) sample of Seyferts, a gradual decline in $\njet \fx / \epsilon$ (and $\njet$) as the system brightens (larger in $\lx/\ledd$) is also observed. Considering the positive dependence on $\lx/\ledd$ in both $\fx$ and $\epsilon$ \citep{VasudevanFabian2009, XieYuan2012}, this implies an efficient reduction in $\njet$, or equivalently a jet quenching behavior, as $\lx/\ledd$ increases. As shown in Figure \ref{fig5}, the whole sample follows a negative relationship
\begin{equation}
\log(\njet\fx / \epsilon)=-0.52^{+0.09}_{-0.09}\log(\lx/\ledd)-2.77^{+0.25}_{-0.25}.\label{eq:nfe_lbol}
\end{equation}
For our adopted values of $\fx$ and $\epsilon$, it can also be expressed as
\begin{equation}
\log \njet = -0.52^{+0.09}_{-0.09} \log(\lbolh/\ledd) - 4.45^{+0.15}_{-0.15},\label{eq:eta_lbol}
\end{equation}
The results of Pearson correlation analysis are provided in Table \ref{tab2}. We note that such a negative correlation is also observed in literature (e.g., \citealt*{Wojtowicz2020}). It is also consistent with the well-known anti-correlation between radio loudness and bolometric luminosity in AGNs (e.g., \citealt*{TerashimaWilson2003, Panessa2007, Ho2008}) and BHBs \citep{Fender2009} as well.

For completeness, we include in Table \ref{tab3} the estimations of the mean values of $\njet$ for different subsamples, i.e. Seyfert 1s vs. Seyfert 2s. 
There is no clear difference in $\njet$ among Sy 1s and Sy 2s. Moreover, although the $\njet$ of RL AGNs is about one order magnitude larger than that of RQ AGNs, it's still smaller than that reported in literature (e.g., \citealt{Inoue2017,Wojtowicz2020}).

Finally we caution that for RQ AGNs, not only the jet power estimation method is uncertain, there are also debates on the radio origin. If other mechanisms play a dominant role, the jet production efficiency reported above will be an over-estimation.

\begin{table}
	\caption{Jet Production Efficiency $\njet$ (and $\njet \fx / \epsilon$) of {\it INTEGRAL}-Selected Seyferts}
	\centering
	\label{tab3}
	\begin{tabular}{rcc}
		\hline
		Sample (No.) 	& ${<}\njet\fx/\epsilon{>}$  	& ${<}\njet{>}$					\\
						& 								& ($\fx=16$ and $\epsilon=8\%$) \\
		\hline
		Full sample (64) 	& $3.8^{+1.7}_{-3.0} \times 10^{-2}$ 	& $1.9^{+0.9}_{-1.5} \times 10^{-4}$ \\
		\hline
		RQ (37) 			& $1.4^{+0.3}_{-0.8} \times 10^{-2}$ 	& $6.9^{+1.7}_{-4.1} \times 10^{-5}$ \\
		\hline
		RL (27) 			& $1.6^{+0.4}_{-0.9} \times 10^{-1}$ 	& $7.9^{+1.9}_{-4.5} \times 10^{-4}$ \\
		\hline
		Sy 1s (35) 			& $2.8^{+1.2}_{-2.2} \times 10^{-2}$ 	& $1.4^{+0.6}_{-1.1} \times 10^{-4}$ \\
		\hline
		Sy 2s (29) 			& $5.5^{+2.5}_{-4.3} \times 10^{-2}$ 	& $2.8^{+1.3}_{-2.2} \times 10^{-4}$ \\
		\hline
	\end{tabular}
\end{table}

\subsection{BHBs versus AGNs}

It is known that there are distinctive differences among bright AGNs and low-luminosity AGNs (e.g., \citealt{Ho2008} for a review), where the separation is around $\lbol/\ledd \simeq (1-2)\%$ or $\lx/\ledd \simeq 10^{-3}$. Direct connections or analogies between bright AGNs and BHBs in soft state (and possibly the intermediate state, see \citealt*{Belloni2010} for state classification in BHBs), and between low-luminosity AGNs and BHBs in hard state, are now crudely established (among others, see e.g., \citealt*{KordingJesterFender2006, Ho2008, YuanNarayan2014, Yang2015}). The accretion modes in the bright AGNs and low-luminosity AGNs correspond to, respectively, the two feedback modes established in AGN feedback field  \citep{Fabian2012, KormendyHo2013, HeckmanBest2014}. The Seyfert galaxies in this work have an X-ray luminosity of $\lx\sim (10^{-4} - 1)\, \ledd$, i.e. most of them belong to the bright AGN regime, thus may be powered by cold accretion disc (e.g., SSD, or the two-phase accretion flow, cf. \citealt*{Yang2015}), and the rest are powered by hot accretion flow. We emphasis that the X-rays cannot be from the SSD, but may be from the corona above (and below) the SSD.

One unresolved puzzle in accretion theory is that, BHBs usually stay in their soft state when $\lx/\ledd \ga 10^{-3}$. In this state, the accretion is a cold SSD \citep{ShakuraSunyaev1973}, and the continuous jet will usually be quenched.\footnote{We note that the compact jet in the soft state is also discovered recently in a limited number of systems and/or outbursts, e.g., Cyg X-3 \citep{Zdziarski2018}, GRS 1739--278 \citep{XieYanWu2020}, and Cyg X-1 \citep{Zdziarski2020}. These observations are of crucial importance, since they find in BHBs the counterparts of RL bright AGNs (and quasars).} On the other hand, although the fraction of radio loudness declines as luminosity increases (e.g., \citealt*{TerashimaWilson2003,Ho2008}, and Sec. \ref{sec:eta_jet}), a significant fraction (typically $\sim 10\%$) of bright AGNs still have detectable radio emission, some of which are even radio loud. Such discrepancy among BHBs and AGNs is fairly robust, but the corresponding physical reason remains unclear (see next section).

\subsection{Jet production}

Theoretically, the jet power is determined by the BH spin and the magnetic flux near BH \citep{BlandfordZnajek1977, Ghisellini2014, Tchekhovskoy2015}. However, in the discrepancy among BHBs and AGNs, the BH spin is known not to be a dominate factor. For example, even for BHBs whose BH spin is large (e.g., $a>0.8$; \citealt*{MillerMiller2015}), the jet is still quenched during soft state. Even for AGNs, it is known that many Seyferts with large BH spin are radio-quiet \citep{Reynolds2014}, i.e. there is no direct link between BH spin and jet power.

The only possible mechanism left relates to the magnetic flux (or somewhat equivalently, the magnetic field strength) near BH. Indeed, theoretically \citet*{HeinzSunyaev2003} argued that there is a $\mbh$-dependence in the jet physics itself. In their scale invariant jet model, a larger $\mbh$ leads to a relatively stronger magnetic field near the BH, which would make it easier to launch a relativistic jet. One possible prediction of this model is that, the jet velocity may be larger in a system with relatively stronger magnetic fields (i.e. larger in $\mbh$), where acceleration may be more efficient. Observationally there is indeed supporting evidence. In AGNs with supermassive BHs ($\mbh\sim 10^{7-9}\msun$) the jets are usually relativistic with $\Gamma\sim 10$ \citep{Kellermann2004}; in narrow-line Seyfert 1s ($\mbh\sim 10^6\msun$), the jets are only mildly relativistic with $\Gamma \sim 3-5$ \citep{Gu2015}; and in the hard state of BHBs ($\mbh\sim 10\msun$) the jet velocity is fairly low with $\Gamma\sim 1-2$ \citep{Fender2009}. 

One important uncertainty in this picture is that, the central engine of bright AGNs as investigated here may be dominated by a cold SSD (thus most prominent in optical and UV radiation; while the X-ray emission is of secondary role), and it remains unclear whether or not a large-scale magnetic field can be developed around a cold disc like SSD. Recent high resolution magnetohydrodynamic simulations of SSDs reveal that, through the surface layer accretion process, the global vertical magnetic fields can be dragged inward and accumulate near BH. Consequently a magnetically-arrested disc (MAD) -- capable of launching a relativistic jet -- can then be achieved (e.g., \citealt{Avara2016, Mishra2020}). However, we emphasis that the dynamical structure of these simulations are highly different to those conventional weakly magnetized ones. Besides, it is unclear why some systems can enter into such an unusual state while the rest cannot. The observations of Seyfert galaxies report a fairly low $\njet$ ($\sim 10^{-4}$; see Figure \ref{fig5} and Equation \ref{eq:eta_lbol}), which is much lower than that predicted by MAD (e.g., \citealt*{Ghisellini2014}). In other words, it may suggest that the central engine of the hard X-ray selected Seyfert galaxies of \citetalias{Panessa2015} is magnetically ``weak'', far away from the highly magnetized MAD state (a similar conclusion, see also \citealt*{Wojtowicz2020} for a sample of young radio galaxies). This is totally different from those $\gamma$-ray blazars as reported in \citet{Ghisellini2014}, where a MAD state with nearly maximal possible jet power is observed.

\subsection{Follow-ups and outlook}

We note that, although the {\it INTEGRAL}/IBIS survey provides a complete sample of Seyfert galaxies from hard X-rays, it is not a deep/sensitive survey. The \citetalias{Panessa2015} sample includes only 79 sources. This limits our capabilities to explore fainter systems. One promising hard X-ray survey is done by the {\it Swift}/BAT \citep{Koss2017}, whose sensitivity is about one order of magnitude higher than the {\it INTEGRAL}/IBIS survey. Investigations based on the {\it Swift}/BAT sample will definitely be helpful to this problem (see e.g., \citealt*{Rusinek2020} for the investigation of $\njet$). Most importantly, the {\it Swift}/BAT survey will enlarge the sample size of sources below $10^{-4}$ in $\lx/\ledd$ and/or less massive in $\mbh$. The first is the regime which is more ideal to the standard so-called ``radiatively-inefficient'' hot accretion (\citealt{YuanNarayan2014}), while the latter will (partially) fullfill the BH mass gap between BHBs and AGNs.

As mentioned in Sec.\ref{sec:rqmodel}, to explore the possible physical models for the radio emission in RQ AGNs, high resolution and high sensitivity radio observations are the keys to unravel the core region. Recently, \citet{Fischer2021} perform simultaneous radio (VLBA) and X-ray observations for 25 AGNs. Despite a low detection rate ($36\%$) in radio, it allows us to explore the possible physical mechanism of RQs. They found AGNs in their sample jump out the FP from VLBA observations, but follow the FP from VLA observations, which implies that this discrepancy might be contributed by extranuclear radio emissions.

\section{Summary and Conclusions}\label{sec:summary}

Sample selection/control is always an important issue in the fundamental plane investigation (e.g., \citealt*{Panessa2007,LiWuWang2008,XieYuan2017,Yao2018}). In this work, we focus on a complete hard X-ray selected sample of Seyfert galaxies originally gathered by \citetalias{Panessa2015}.  We only include sources that have measurements/estimations of BH mass and luminosities in radio and X-rays. Our final near-complete sample includes 64 (out of the original 79) sources, among which 35 are Seyfert 1s and 29 are Seyferts 2s, or 27 are RL and 37 are RQ. The dynamical range in $\lx/\ledd$ is between $\sim10^{-4}$ and $\sim10^{-0.5}$. According to the typical separation of $\lx/\ledd \sim 10^{-3}$ (e.g., \citealt*{Ho2008, Yang2015}), most sources of our sample belong to bright AGNs, and a small fraction are low-luminosity ones.

This near-complete hard X-ray selected sample suffers a strong dependence of $\lx/\ledd$ on $\mbh$, as clearly shown in the left panel of Figure \ref{fig1}. We thus limit ourselves only to probe the radio/X-ray correlation in Eddington units (e.g., $\lr/\ledd$ vs. $\lx/\ledd$). This choice reflects our motivation to investigate the accretion theory, where only luminosities in Eddington units are related to the accretion mode (see \citealt*{Esin1997,Ho2008,YuanNarayan2014}). 

Our main results can be summarized as follows.
\begin{itemize}
	\item There is no clear difference in $\lhx/\lx$ (mean value, scatter, distribution) among Seyfert 1s and Seyfert 2s, and among RL Seyferts and RQ Seyferts. This may suggest a common origin of X-rays in our sample.
	\item  The slope of the radio/X-ray correlation is almost universal among different types of sources, i.e. it is irrelevant to the broad emission line properties (Seyfert 1s or 2s), the radio loudness (RL or RQ), and the morphological size of radio emission (extended or compact). The whole sample follows $\lr/\ledd \propto (\lx/\ledd)^{0.77\pm0.10}$, which agrees with the standard FP at $2\sigma$ level \citep{Merloni2003}.
	\item Under the jet origin interpretation, the average jet production efficiency of Seyfert galaxies is ${<}\njet{>}=1.9^{+0.9}_{-1.5} \times 10^{-4}$, which is two orders of magnitude lower than that of RL sources at even higher bolometric luminosities \citep{vanVelzenFalcke2013, Inoue2017,SoaresNemmen2020, Wojtowicz2020}. More specificaly, we have ${<}\njet{>}=7.9^{+1.9}_{-4.5} \times 10^{-4}$ for RL Seyferts and ${<}\njet{>}=6.9^{+1.7}_{-4.1} \times 10^{-5}$ for RQ Seyferts. Besides, a gradual decline in $\njet$ as the system brightens is also evident, i.e. $\njet \propto (L_{\rm bol,h}/\ledd)^{-0.52\pm0.09}$. Such jet quenching process during the brightening phase is expected in both theory and observations. 
\end{itemize}

\section*{Acknowledgments}

We appreciate the referee for his/her detailed report and thoughtful comments (especially updates of radio emission in RQ systems) that improve our presentation. This work was supported by the National Key R\&D Program of China (NKRDC, 2018YFA0404602), and the Key Laboratory of Radio Astronomy, Chinese Academy of Sciences (CAS). FGX was supported in part by National SKA Project of China No. 2020SKA0110102, the National Science Foundation of China (NSFC, 11873074) and the Youth Innovation Promotion Association of CAS. LCH was supported by NKRDC (2016YFA0400702) and NSFC (11721303 and 11991052). This work has made extensive use of the NASA/IPAC Extragalactic Database (NED), which is operated by the Jet Propulsion Laboratory, California Institute of Technology, under contract with the NASA. This research has made use of the SIMBAD database, operated at CDS, Strasbourg, France.

%

\bsp    
\label{lastpage}
\end{document}